\documentclass[1p]{elsarticle}

\usepackage{times}
\usepackage[colorlinks=true]{hyperref}

\usepackage[version=4]{mhchem}
\usepackage{graphicx}

\usepackage{soul}
\usepackage{color}
\sethlcolor{yellow}

\usepackage{booktabs}
\usepackage{multirow}

\usepackage{fancyhdr}
\makeatletter
\def\ps@pprintTitle{%
	\let\@oddhead\@empty
	\let\@evenhead\@empty
	\let\@oddfoot\@empty
	\let\@evenfoot\@empty}
\makeatother

\newcommand{\cell}[4]{\setlength{\tabcolsep}{2pt}\begin{tabular}{lr}TP:&#1\%\\TN:&#2\%\\FP:&#3\%\\FN:&#4\%\end{tabular}}

\begin{document}

\begin{frontmatter}

\title{Applications of Deep Learning to Nuclear Fusion Research\tnoteref{t1}}
\tnotetext[t1]{This paper is based on a talk presented at NVIDIA's GPU Technology Conference, which took place at the International Congress Center in Munich, Germany, October 9--11, 2018 (GTC Europe 2018).}

\author[]{Diogo R. Ferreira$^{\text{1}}$}

\author[]{JET Contributors\corref{jet}}
\cortext[jet]{See the author list of X. Litaudon et al 2017 Nucl. Fusion 57 102001}

\renewcommand{\elsaddress}{EUROfusion Consortium, JET, Culham Science Centre, Abingdon, OX14 3DB, UK\\
                           $^{\text{1}}$Instituto de Plasmas e Fus\~{a}o Nuclear, Instituto Superior T\'{e}cnico, Universidade de Lisboa, 1049-001 Lisboa, Portugal}



\begin{abstract}
Nuclear fusion is the process that powers the sun, and it is one of the best hopes to achieve a virtually unlimited energy source for the future of humanity. However, reproducing sustainable nuclear fusion reactions here on Earth is a tremendous scientific and technical challenge. Special devices -- called tokamaks -- have been built around the world, with JET (Joint European Torus, in the UK) being the largest tokamak currently in operation. Such devices confine matter and heat it up to extremely high temperatures, creating a plasma where fusion reactions begin to occur. JET has over one hundred diagnostic systems to monitor what happens inside the plasma, and each 30-second experiment (or pulse) generates about 50 GB of data. In this work, we show how convolutional neural networks (CNNs) can be used to reconstruct the 2D plasma profile inside the device based on data coming from those diagnostics. We also discuss how recurrent neural networks (RNNs) can be used to predict plasma disruptions, which are one of the major problems affecting tokamaks today. Training of such networks is done on NVIDIA GPUs.
\end{abstract}

\end{frontmatter}

\section{Introduction}

The Sun is the star at the center of our solar system, and nuclear fusion is the process that powers the Sun. At its hot and dense core, the Sun fuses hydrogen into helium, and since the result has less mass that the original elements, the excess mass is released as energy.

The idea of replicating this process on Earth (i.e.~of using sustainable fusion reactions as a source of energy) dates back to at least the mid-20th century. For practical reasons, one of the most promising reactions is the one involving two isotopes of hydrogen:
\begin{equation}
\ce{^2_1H + ^3_1H -> ^4_2He + ^1_0n + 17.6MeV}
\end{equation}
where $\ce{^2_1H}$ (\emph{deuterium}) and $\ce{^3_1H}$ (\emph{tritium}) fuse into a helium nucleus, releasing a neutron and a relatively large amount of energy, when compared to other possible fusion reactions~\cite{chen16intro}.

However, the engineering a fusion reactor has met some technical difficulties, the most important of which being how to successfully confine matter in a hot, plasma state. In the Sun, such problem is solved by the gravitational field, which holds the plasma together. On Earth, keeping the plasma away from any other material (which would eventually cool it down) has turned out to be a surprisingly difficult challenge.

Several different types of fusion reactors have been devised based on the idea of confining the plasma with magnetic fields~\cite{freidberg14ideal}. One of the most successful designs is the \emph{tokamak}~\cite{lackner12tokamaks}, which confines the plasma in a toroidal shape through the use of a toroidal field together with a poloidal field to yield an overall helical field around the torus.

At the macroscopic level, a plasma is approximately neutral (a concept that is referred to as \emph{quasi-neutrality}~\cite{bellan06fundamentals}), but the fact that electrons and nuclei get stripped from each other at very high temperatures allows for the treatment of a plasma as a two-fluid model~\cite{goedbloed04principles}. The two fluids (of positively and negatively charged particles) can be confined by magnetic fields.

The largest tokamak currently in operation is the Joint European Torus (JET) located near Oxford, in the UK. JET is a D-shaped tokamak with a major radius of 2.96 m and a minor radius in the range 1.25--2.10 m~\cite{horton16joint}. A larger machine with a similar design -- the International Thermonuclear Experimental Reactor (ITER) -- is currently being built in Cadarache, France.

JET has a vast assortment of diagnostic systems, including magnetic coils to measure plasma current and instabilities, interferometers and reflectometers to measure plasma density, Thomson scattering to determine the electron temperature, spectroscopy to measure ion temperature, and X-ray cameras to measure electromagnetic radiation, among many others~\cite{wesson11tokamaks}.

One of the diagnostic systems that we will be focusing on -- in fact, the only diagnostic that we will be using in this work -- is the bolometer system available at JET~\cite{huber07upgraded}. This bolometer system measures the plasma radiation on a cross-section of the fusion device.

\begin{figure}[h]
	\centering
	\includegraphics[scale=0.5]{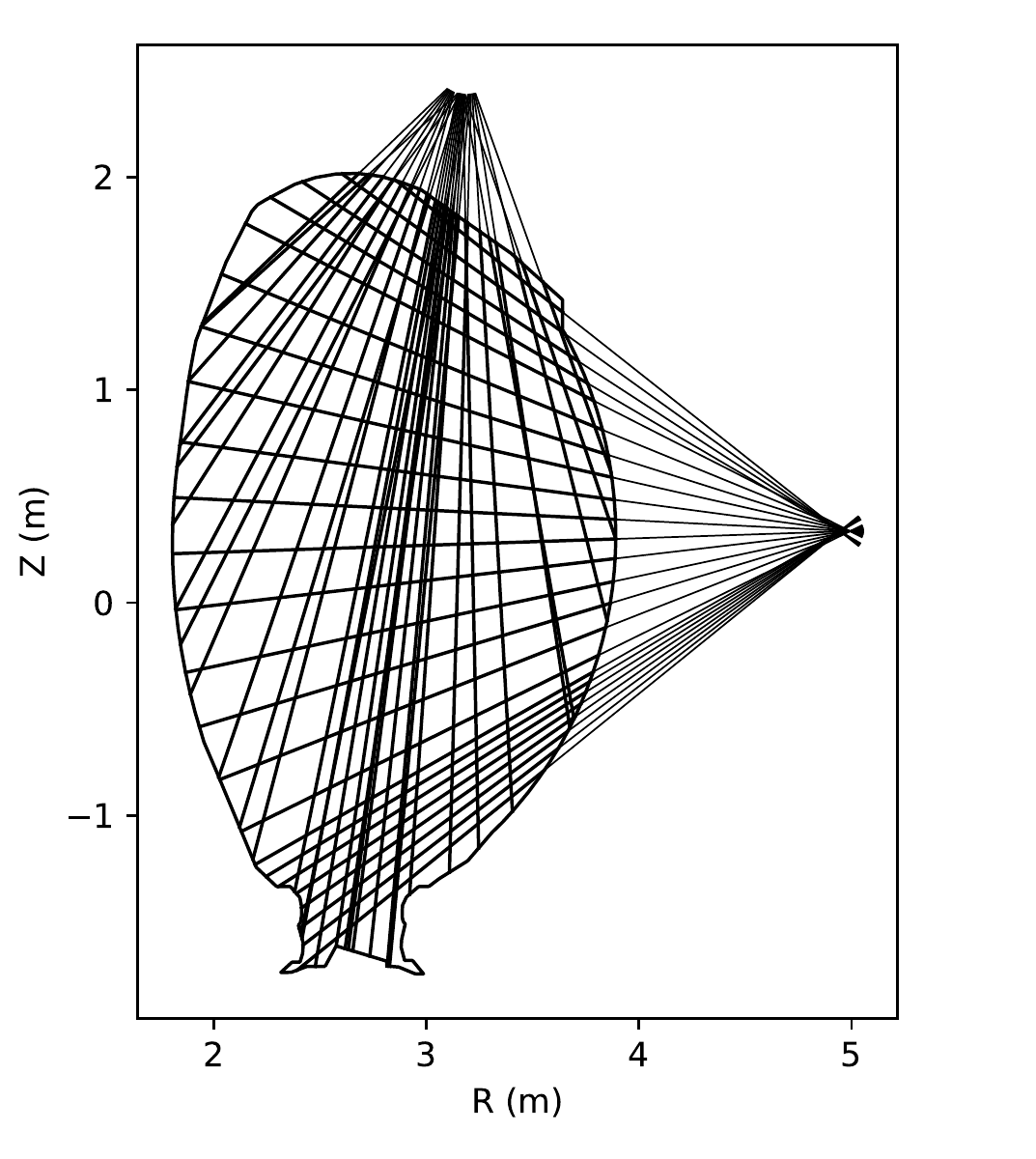}
	\caption{Lines of sight for the vertical and horizontal cameras of the bolometer system at JET.}
	\label{fig_kb5}
\end{figure}

The system is illustrated in Figure~\ref{fig_kb5}. It comprises a horizontal camera and a vertical camera with multiple lines of sight. Some of these lines of sight are directed towards a region of special interest called the \emph{divertor}~\cite{matthews09development}, where a lot of activity takes place due to plasma exhaust.

The diagnostic is called a \emph{bolometer system} because each camera comprises an array of sensors called \emph{bolometers}. Each bolometer measures the radiation along a particular line of sight (technically, it provides a measure of the line-integrated radiation). In essence, a bolometer is a thin metal foil attached to a temperature-sensitive resistor~\cite{ingesson08tomography}. As the metal foil absorbs radiation, its temperature changes and there is a change in resistance. This provides an effective way to measure electromagnetic radiation in the range from ultraviolets (UV) to soft X-rays.

Using both the horizontal and the vertical camera, it is possible to reconstruct the 2D plasma radiation profile through a process called \emph{tomography}~\cite{ingesson98softxray}. The underlying principle is the same as Computed Tomography (CT) in medical applications~\cite{herman09fundamentals}, but the reconstruction technique is different due to the relatively scarce number of lines of sight available.

\begin{figure}[h]
	\centering
	\includegraphics[scale=0.4]{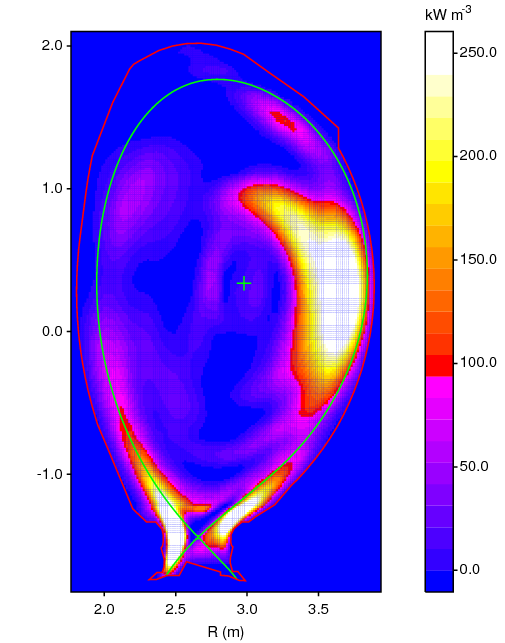}
	\caption{Sample tomographic reconstruction for pulse 89065 at $t$ = 47.8 s.}
	\label{fig_tomo}
\end{figure}

Figure~\ref{fig_tomo} shows a sample tomographic reconstruction computed at JET. The reconstruction method involves an iterative constrained optimization approach that minimizes the error with respect to the observed measurements, while requiring the solution to be non-negative~\cite{fehmers98algorithm}.

This iterative reconstruction method takes a significant amount of computation time. The total run-time
depends on the actual data but, with the code available at JET, it can take up to one hour to produce a single reconstruction. 

To appreciate the computational effort involved, one should consider the following:
\begin{itemize}

	\item The bolometer system has a sampling rate of 5 kHz.

	\item For the purpose of noise reduction, a window average of 5 ms is usually applied, which corresponds to 25 samples.
	
	\item Subsampling by a factor of 25 yields an effective sampling rate of 200 Hz.
	
	\item So, in principle, it should be possible to have 200 reconstructions per second.
	
	\item For a typical 30-second pulse at JET, this would mean a total of 6000 reconstructions.
	
	\item At one hour per reconstruction, this could take up to 250 days of computation time.

\end{itemize}

Clearly, another way to compute the reconstructions for an entire pulse should be found. The following section describes
how deep learning was applied to train a convolutional neural network (CNN) that is able to compute 3000 reconstructions per second on an NVIDIA GPU~\cite{ferreira18fullpulse}.

\section{Deep Learning for Plasma Tomography}

Deep learning~\cite{lecun15deep} has had a tremendous impact in several fields, such as image processing and natural language processing. In particular, convolutional neural networks (CNNs) have been very successful at classifying input images into a set of output classes. This has been demonstrated in the recognition of hand-written digits~\cite{lecun98gradient}, in the classification of Web images~\cite{krizhevsky12imagenet}, and in the annotation of online videos~\cite{karpathy14video}, to cite only a few examples.

In general, CNNs have a structure that comprises one or more convolutional layers. Each convolutional layer applies multiple filters (in the form of a small kernel or sliding window) to the input. The purpose of a filter is to detect a specific feature, so its output is called a \emph{feature map}. In a CNN, the first convolutional layer operates directly on the input image; subsequent layers operate on the feature maps produced by previous layers.

In addition to convolutional layers, CNNs contain subsampling layers. There is usually one subsampling layer after each convolutional layer. The purpose of having subsampling is to make the feature maps smaller and allow their number to progressively increase after each convolution, while keeping the network under a manageable size.

After the convolutional and subsampling layers, the end result is usually a large number of small feature maps. These are then flattened and connected to a couple of dense layers, which perform the actual classification. The main idea behind a CNN is to have a first stage of convolutional layers to extract meaningful features from the input image, and a second stage of dense layers to perform classification based on those features.

In summary, the input to a CNN is typically a 2D image and its output is a 1D vector of class probabilities. However, for the purpose of plasma tomography, it would be useful to have a network that takes a 1D vector of bolometer measurements as input, and produces a 2D image of the plasma radiation profile as output.

In the literature, the inverse of a CNN has been referred to as a \emph{deconvolutional network}, and it has found applications in image segmentation~\cite{noh15learning} and object generation~\cite{dosovitskiy15learning}. For example, by specifying the class label and the camera position (1D data), a deconvolutional network is able to generate an object (2D image) of the specified class from the given camera view~\cite{dosovitskiy17learning}.

\begin{figure}[h]
	\centering
	\includegraphics[scale=0.5]{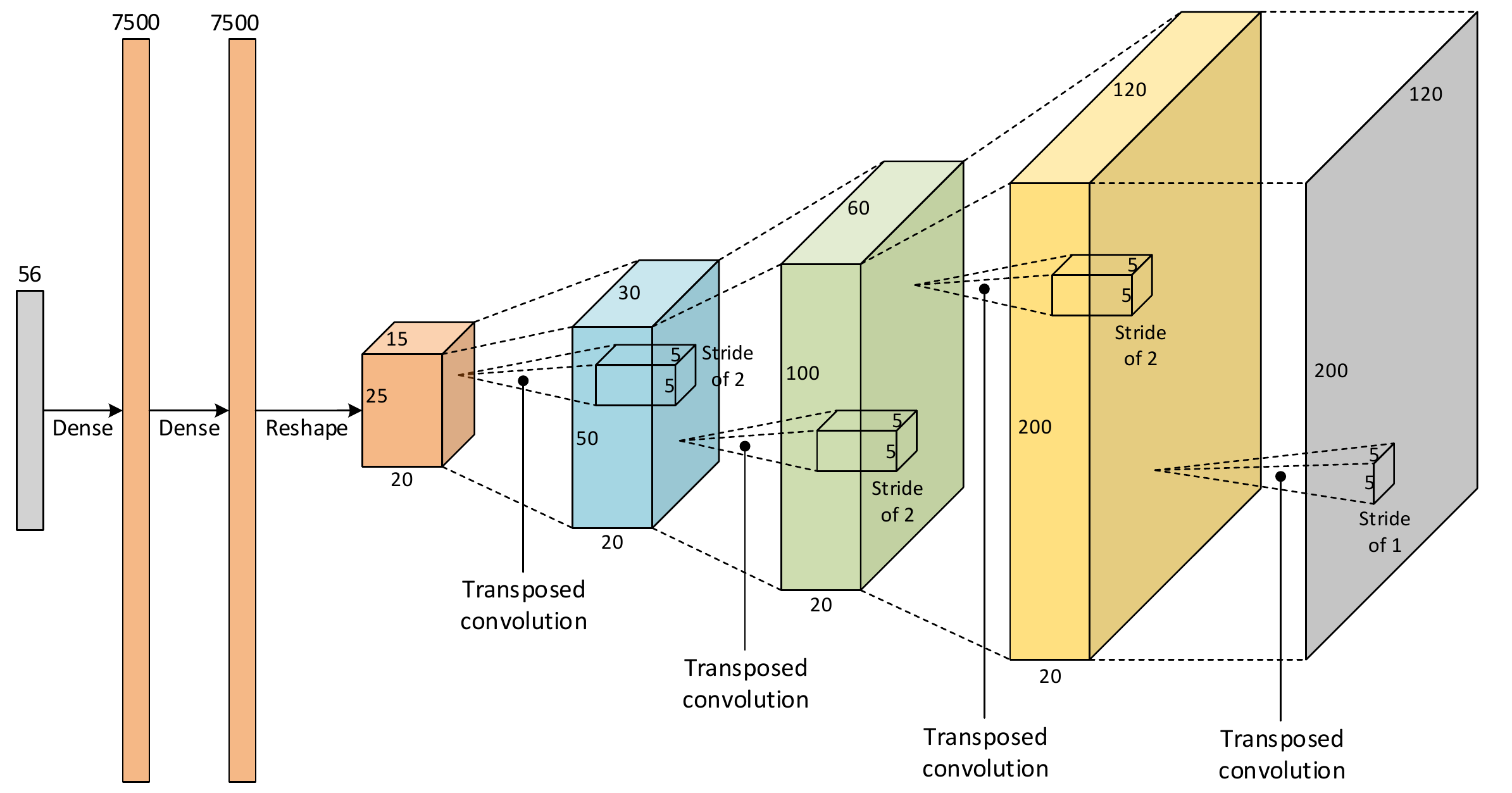}
	\caption{Deconvolutional neural network for plasma tomography.}
	\label{fig_cnn}
\end{figure}

In this work, we use a deconvolutional network that receives the bolometer measurement as input (a total of 56 lines of sight from both cameras) and produces a 120$\times$200 reconstruction of the plasma radiation profile. The network architecture is shown in Figure~\ref{fig_cnn}.

At the beginning, there are two dense layers with 7500 nodes, which can be reshaped into a 3D structure of size 25$\times$15$\times$20. This structure can be regarded as comprising 20 features maps of size 25$\times$15. By applying a series of transposed convolutions, these features maps are brought up to a size of 200$\times$120, from which the output image is generated by one last convolution.

Essentially, the \emph{transposed convolution} is the inverse of the convolution in the sense that, if a sliding window would be applied to the output, the result would be the feature map given as input. It can be shown that learning a transposed convolution is equivalent to learning a weight matrix that is the transpose of a regular convolution, hence the name of this operation~\cite{dumoulin16guide}.

In Figure~\ref{fig_cnn}, the upsampling of feature maps from 25$\times$15 up to 200$\times$120 is achieved by having each transposed convolution operate with a stride of two pixels (i.e.~one pixel is being skipped between each two consecutive positions of the sliding window). This means that the output is two times larger and taller than the input, except for the very last convolution which uses a stride of one to keep the same size.

To train the network shown in Figure~\ref{fig_cnn}, we gathered the tomographic reconstructions that have been computed at JET for all the experimental campaigns between 2011 and 2016. This yielded a total of 25584 reconstructions, which have been divided into 90\% (23025) for training, and 10\% (2559) for validation.

The network was trained on an NVIDIA Tesla P100 GPU using accelerated gradient descent (Adam~\cite{kingma14adam}) with a small learning rate ($\text{10}^{-\text{4}}$) and a batch size of 307 samples. The batch size was chosen to be a perfect divisor of the number of training samples (23025/307 = 75) in order to avoid having any partially filled batch in the training set.

\begin{figure}[h]
	\centering
	\includegraphics[scale=0.5]{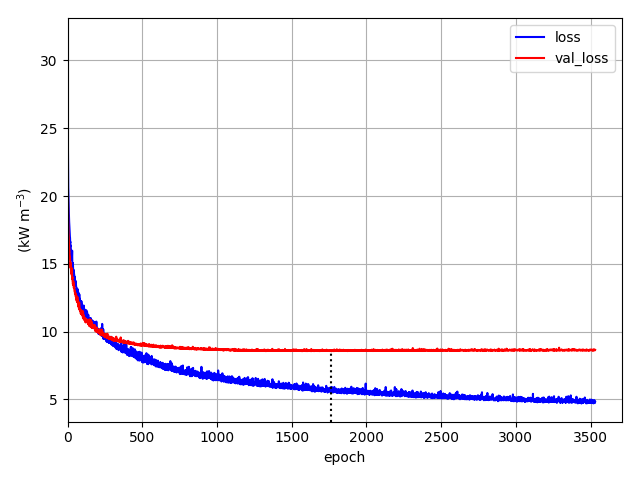}
	\caption{Loss and validation loss during training of the deconvolutional network.}
	\label{fig_loss_cnn}
\end{figure}

Figure~\ref{fig_loss_cnn} shows the evolution of the training loss and of the validation loss across 3530 epochs, which took about 20 hours. The best result was achieved at epoch 1765, with a minimum validation loss of 8.556 kW m$^{-\text{3}}$. This is the mean absolute error per pixel, which can be compared to the dynamic range of Figure~\ref{fig_tomo} (8.556/250 $\approx$ 3.4\% error).

Once trained, the network can be used to generate the reconstruction for any given pulse at any given point in time. In fact, given the bolometer data from an entire pulse, the network can generate the reconstructions for all points in time. This way, it becomes possible to analyze the evolution of the plasma radiation profile across the entire experiment.

\begin{figure}[h]
	\centering
	\includegraphics[width=\textwidth]{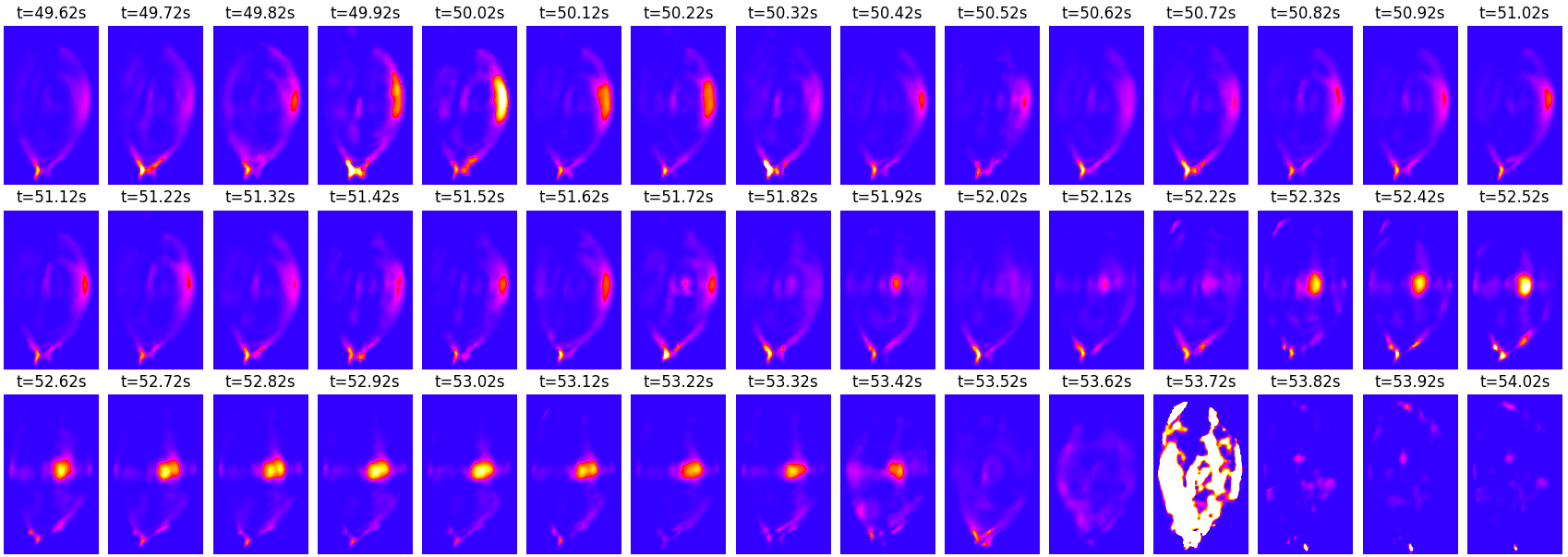}
	\caption{Reconstruction of pulse 92213 from $t$ = 49.62 s to $t$ = 54.02 s with a time step of 0.1 s.}
	\label{fig_pulse}
\end{figure}

Figure~\ref{fig_pulse} shows the reconstruction of pulse 92213 from $t$ = 49.62 s onwards, with a time increment of 0.1 s (due to space restrictions only). The first row shows a focus of radiation developing on the outer wall, which seems to slowly fade away (rows 1--2), only to reappear later at the plasma core with particularly strong intensity (rows 2--3).

The radiation peaking stays at the core for a relatively long time (at least from $t$~=~52.32 s to $t$~=~53.42~s) while changing slightly in shape during that interval. Eventually, it also fades away as the heating systems are being turned off. However, just as it seemed that the plasma was about to soft land, there is a disruption at $t$ = 53.72 s.

Disruptions are one of the major problems affecting tokamaks, and they are one of the major impediments towards fusion energy today. The dynamics of disruptions are not yet fully understood, but one of the main causes seems to be impurity accumulation at the plasma core, which decreases the core temperature and eventually leads to core collapse~\cite{vries11survey}.

To some extent, this phenomenon is clearly visible in Figure~\ref{fig_pulse}. It is therefore not surprising that the bolometer system has played an key role in several disruption studies~\cite{arnoux09heat,riccardo10disruption,huber11radiation,lehnen11disruption}, either by providing a measurement of total radiation or by providing the tomographic reconstruction at a few, specific points in time. With the proposed deconvolutional network, it is possible to quickly obtain the plasma radiation profile for all points in time.

\section{Deep Learning for Disruption Prediction}

Over the years, a number of different approaches to disruption prediction have been developed, including the use of neural networks~\cite{pautasso02online,cannas04disruption,windsor05cross,yoshino05neural}, support vector machines~\cite{cannas07svm,ratta10advanced,lopez14implementation} and decision trees~\cite{murari08prototype,murari09unbiased}, among other techniques. In all of these approaches, the prediction model takes a set of global plasma parameters as input (such as the plasma current, the locked mode amplitude, the total input power, the safety factor, the poloidal beta, etc.) and outputs a probability of disruption, or the remaining time to a predicted disruption.

More recently, the use of random forests~\cite{rea18exploratory} and even deep learning~\cite{svyatkovskiy17training} have shown promising results, but the input signals that are provided to these models still consist of about ten global plasma parameters derived from different diagnostics.

In the previous section, we have seen that a bolometer system with multiple lines of sight is able to capture some of the physical phenomena associated with plasma disruptions. From such bolometer data, it is possible to reconstruct the internal shape of the plasma radiation profile and observe, for example, an impurity concentration developing at the plasma core, which could be the precursor of an impending disruption.

This suggests that the array of sensor measurements coming the from multiple cameras of a bolometer system could, in principle, be used for disruption prediction. The main difference to previous works is that the prediction model will be working on the basis of a diagnostic with multiple views into the plasma, rather than on a set of global plasma parameters.

In addition to having a spatial view into the plasma, another dimension that will be important to consider is time. As suggested by Figure~\ref{fig_pulse}, the phenomenona that precede a disruption will develop and endure for some time, on the order of seconds. Therefore, it will be important to consider not only the bolometer measurements at a certain point in time, but also the preceding measurements in order to have a sense of how the plasma behavior is developing (e.g.~to determine whether it is moving closer to, or away from, a possible disruption).

With both dimensions, the input to the predictor will be sequence of vectors, where each vector contains the bolometer measurements from the multiple lines of sight for a given point in time. This closely resembles the way in which recurrent neural networks are used in natural language processing (NLP), where each word in a sentence (the sequence) is mapped into a vector space of a certain dimensionality, through a process called \emph{word embedding}~\cite{goldberg17neural}.

A recurrent neural network (RNN) is able to handle sequences of vectors by processing each vector at a time. The key distinctive feature of RNNs is that they maintain an internal state, and this internal state is fed back to the input along with the next vector in the sequence. A RNN can therefore ``remember'' features from past vectors in the sequence.

In simple RNNs, such ``memory'' is necessarily short-lived, since the network keeps updating its internal state as it receives new vectors, and gradually ``forgets'' about vectors that are in the distant past. To provide RNNs with the ability to maintain long-term dependencies, more advanced architectures have been proposed, namely the \emph{long short-term memory} (LSTM)~\cite{hochreiter97lstm,gers02learning}. In an LSTM, information can be carried across multiple time steps, and that information can be updated or forcefully forgotten at each time step, through the use of an \emph{input gate} and a \emph{forget gate} respectively, while also maintaining an internal state.

More recently, the combination of CNNs and RNNs has been shown to be very effective in tasks such as video processing~\cite{donahue15longterm,donahue17longterm} and text classification~\cite{wang16dimensional,wang16coling}. In essence, the CNN is used as a feature extractor before passing the data to a RNN for sequential processing.

In this work, we use a similar approach to preprocess the bolometer data through a series of 1D convolutional layers before handing them over to an LSTM. Besides feature extraction, this has the advantage of further reducing the noise in the bolometer data. As in a traditional CNN, we use subsampling after each convolutional layer to progressively reduce the length of the time sequence, while increasing the number of filters that are applied to it.

\begin{figure}[h]
	\centering
	\includegraphics[width=\textwidth]{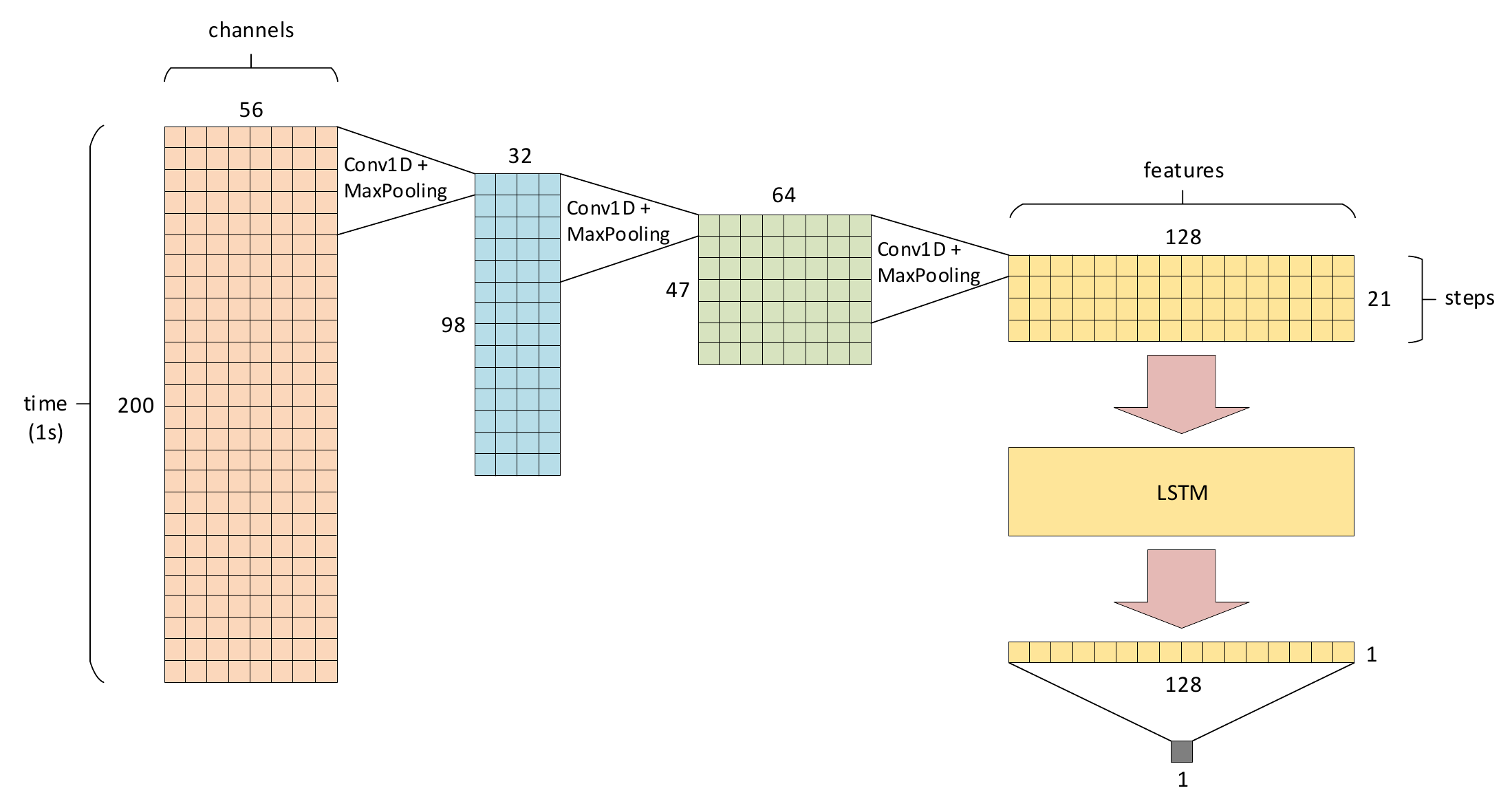}
	\caption{Recurrent neural network for disruption prediction.}
	\label{fig_rnn}
\end{figure}

The resulting network is shown in Figure~\ref{fig_rnn}. The input has a shape of 200$\times$56, with the first being the sequence length, and the latter being the vector size. We use sequences of 200 time points, which corresponds to the 200 Hz sampling rate of the bolometer system, meaning that a full second of bolometer measurements is analyzed by the network in order to produce a prediction. The vector size of 56 corresponds to the number of lines of sight available.

Some of these lines of sight yield always a measurement of zero, since they are reserve channels that typically remain unused. A few other lines of sight are associated with bolometers that are known to be faulty, providing erratic measurements. In these cases, the network is expected to filter out those channels, so the number of filters used in the first convolutional layer (32) is actually less than the number of input channels (56).

In subsequent layers, the number of filters doubles after each convolution. A kernel size of 5 time steps is used in all convolutions, together with a max-pooling operation in order to reduce the sequence length to half. Eventually, the convolutional part of the network yields a sequence of 21 steps, where each step contains a 128-feature vector.

This sequence of feature vectors is handed over to an LSTM with 128 units, which returns the last output after having processed the whole sequence. One final dense layer combines these vector elements into a single output, which is the network prediction.

As mentioned earlier, there are two ways of predicting disruptions:
\begin{itemize}

	\item One way is to predict the \emph{probability of disruption}, i.e.~the probability that the current pulse will end in a disruption. For this purpose, the last layer of the network will contain a sigmoid activation, to produce a value between 0 and 1. The loss function used to train the network will be the binary cross-entropy~\cite{goodfellow16book}, and the network will be trained with sample data from both disruptive and non-disruptive pulses.
	
	\item Another way is to predict the \emph{time to disruption}, i.e.~the remaining time up to a incoming disruption. For this purpose, the last layer of the network will have a linear activation (i.e.~no activation) so that it can produce positive as well as negative values within any range. The loss function used to train the network is the mean absolute error, and the network is trained with sample data from disruptive pulses only.

\end{itemize}

To train these two variants, we gathered the plasma current and the bolometer data for all pulses in the experimental campaigns from 2011 to 2016. The time of disruption was defined as the point in time where the current gradient exceeded 20 MA/s, as is standard practice at JET~\cite{morris12doc}. From a universe of 9798 pulses in those experimental campaigns, only 1683 pulses (about 17\%) reached such threshold and were therefore considered to be disruptive. Hence:
\begin{itemize}

	\item To train the network for probability of disruption, we divided the 9798 pulses into 90\% for training and 10\% for validation. We used a data generator to draw random samples from each pulse and feed them to the network, together with a binary label to indicate whether the sample was drawn from a disruptive or non-disruptive pulse.
	
	\item To train the network for time to disruption, we divided the 1683 disruptive pulses into 90\% for training and 10\% for validation. We used a data generator to draw random samples from each pulse and feed them to the network, together with a label to indicate the timespan between the sample time and the disruption time for that pulse.

\end{itemize}

In both cases, a batch size of 2000 samples was used, where each sample is a 200$\times$56 sequence of vectors, and the sample time is defined as the latest time point in the sequence. Each training epoch was defined as a run through 100 batches. For validation purposes, a similar data generator was used to draw random samples from the validation pulses, with the same batch size, but using only 10 batches for validation at the end of each training epoch.

\begin{figure}[b]
	\centering
	\begin{tabular}{cc}
	\includegraphics[scale=0.4]{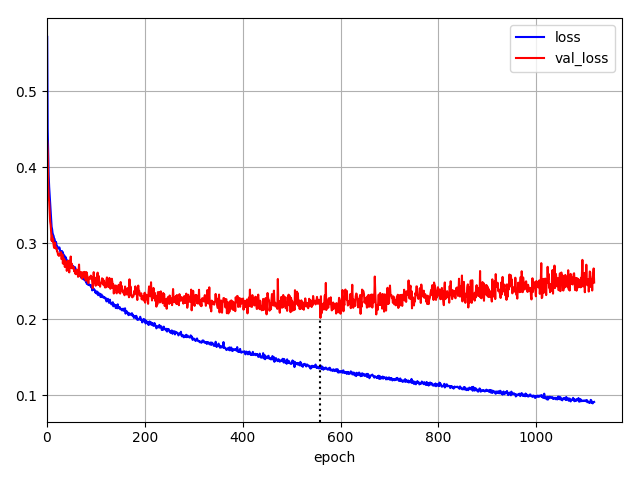} &
	\includegraphics[scale=0.4]{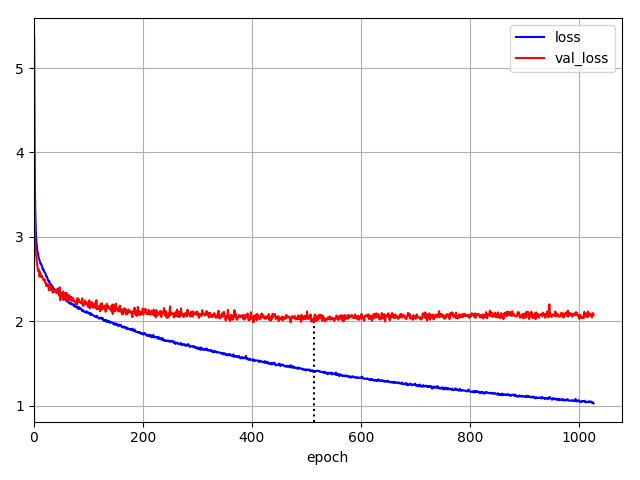}
	\end{tabular}
	\caption{Loss and validation loss for probability of disruption (\emph{left}) and time to disruption (\emph{right}).}
	\label{fig_loss_rnn}
\end{figure}

Figure~\ref{fig_loss_rnn} shows the evolution of the training loss and of the validation loss for both variants:
\begin{itemize}

	\item For probability of disruption, training ran for a total of 1118 epochs (14 hours on an NVIDIA Tesla P100 GPU), with the best result being achieved at epoch 559 with a validation loss (binary cross-entropy) of 0.202.

	\item  For time to disruption, training ran for 1026 epochs (13 hours on the same GPU) with the best result being achieved at epoch 513 with a validation loss (mean absolute error) of 1.986 seconds.

\end{itemize}

Although these loss values appear to be relatively high, it should be noted that they are averaged over all samples used for training/validation. For samples that are far away from the disruption, their loss value will be quite high, because it is difficult to predict when or if a disruption will occur. This will be compensated by a small loss value for samples that are near the disruption, when it will be more important to have a correct prediction.

\begin{figure}[h]
	\centering
	\includegraphics[scale=0.32]{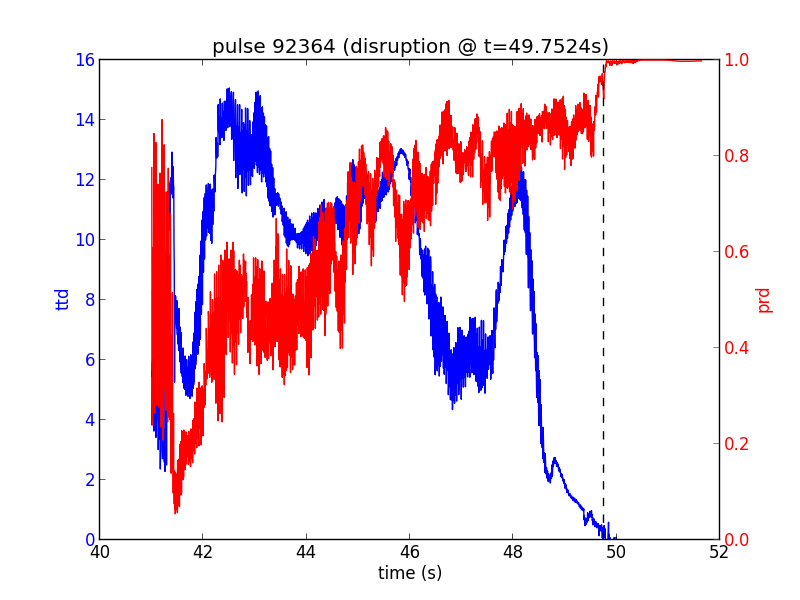}
	\caption{Time to disruption and probability of disruption for pulse 92364.}
	\label{fig_92364}
\end{figure}

Figure~\ref{fig_92364} shows an example of how the predicted time to disruption (\emph{ttd}) and the predicted probability of disruption (\emph{prd}) evolve across an entire pulse. There is an upward trend in the probability of disruption and a downward trend in the time to disruption. Eventually, as the pulse gets close to the disruption, the probability gets close to 1.0 and the time to disruption approaches zero, indicating that a disruption is imminent.

As soon as the probability of disruption exceeds a certain threshold \emph{and} the time to disruption falls below a certain threshold, it is possible to raise an alarm that will trigger preventive actions, such as activating the \emph{disruption mitigation valve} (DMV)~\cite{lehnen11disruption} which, in a matter of a few milliseconds, will inject a massive amount of gas to cool down the plasma.

\begin{figure}[h]
	\centering
	\begin{tabular}{cc}
		\includegraphics[scale=0.32]{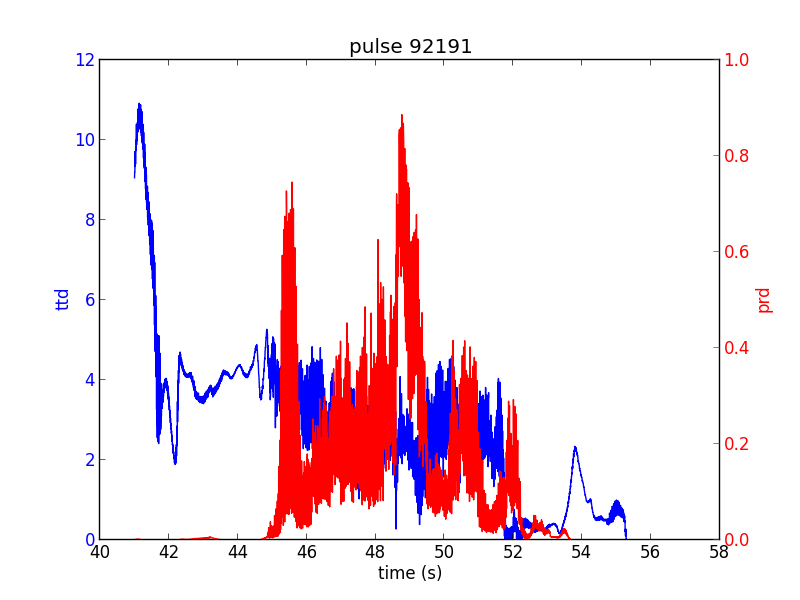} &
		\includegraphics[scale=0.32]{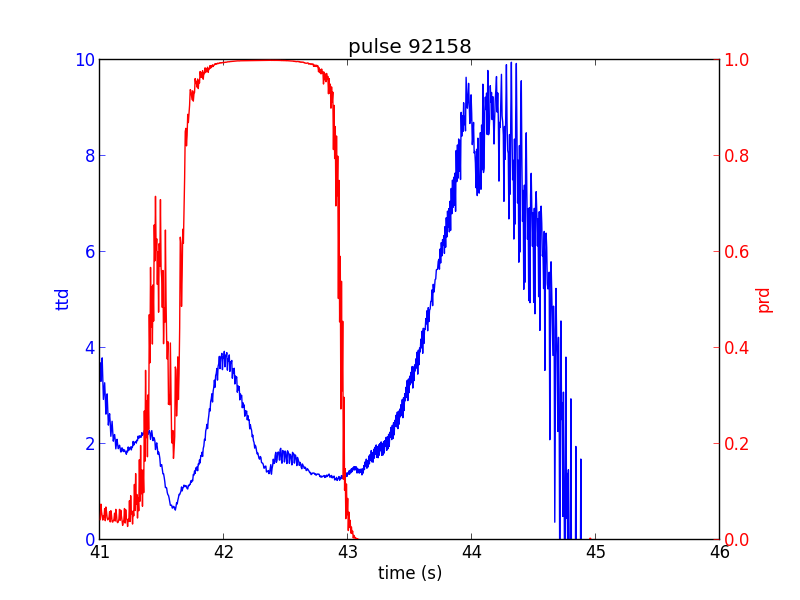}
	\end{tabular}
	\caption{Time to disruption and probability of disruption for pulses 92191 (\emph{left}) and 92158 (\emph{right}).}
	\label{fig_92191_92158}
\end{figure}

Such alarm should not be triggered when the time to disruption is low but the probability of disruption is also low, or when the probability of disruption is high but the time to disruption is also high, as illustrated in the examples of Figure~\ref{fig_92191_92158}. Both of these pulses are non-disruptive.

\begin{figure}[h]
	\centering
	\begin{tabular}{cc}
		\includegraphics[scale=0.32]{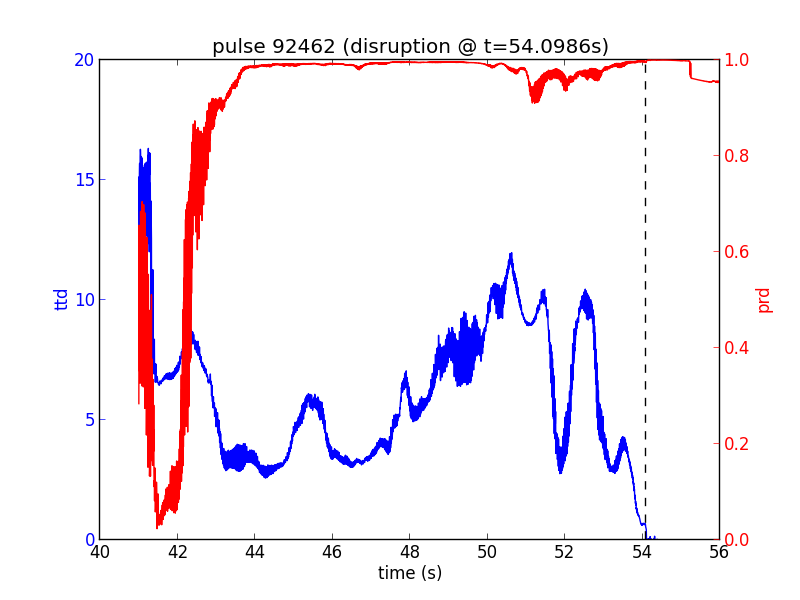} &
		\includegraphics[scale=0.32]{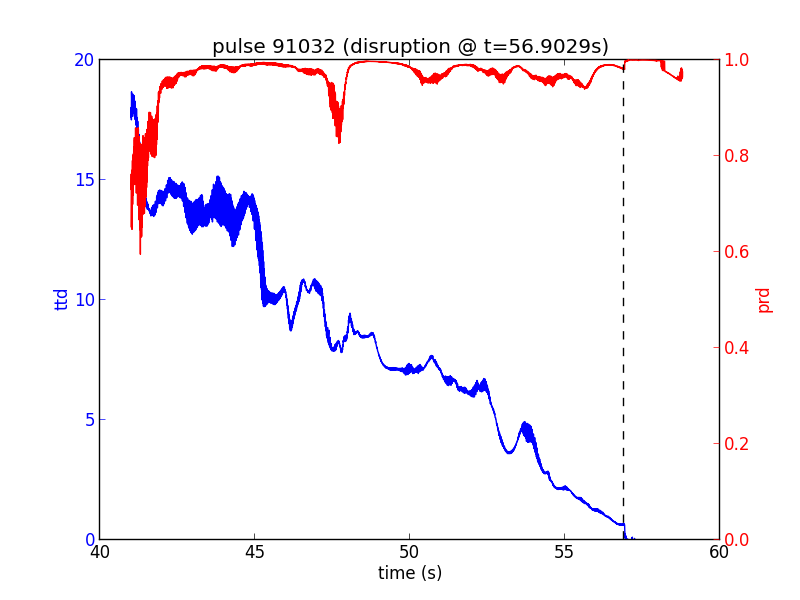}
	\end{tabular}
	\caption{Time to disruption and probability of disruption for pulses 92462 (\emph{left}) and 91032 (\emph{right}).}
	\label{fig_92462_92158}
\end{figure}

A surprising result from this approach is that, for several disruptive pulses, the probability of disruption gets close to 1 early on, and stays there for almost the entire pulse until a disruption eventually occurs. Figure~\ref{fig_92462_92158} shows two examples. Of course, an alarm would only be triggered when the time to disruption also approaches zero, but it is still puzzling how the network was able to recognize signs of disruption so early in the pulse. We hypothesize that the network has learned that certain radiation patterns are always associated with disruptive pulses, even if those radiation patterns are not related to the actual cause of disruption. 

\begin{figure}[h]
	\centering
	\includegraphics[scale=0.32]{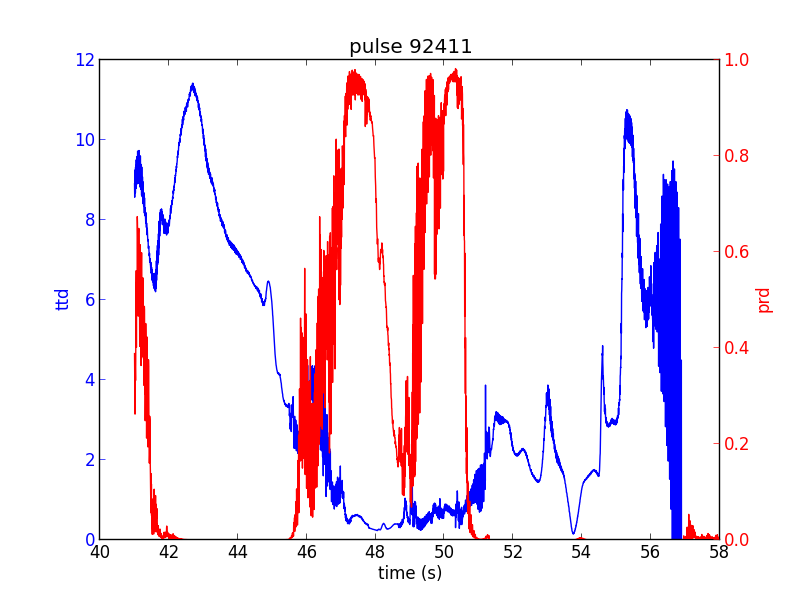}
	\caption{Time to disruption and probability of disruption for pulse 92411.}
	\label{fig_92411}
\end{figure}

In Figure~\ref{fig_92411}, we show an example of how false alarms may occur. During the pulse, the probability of disruption becomes very high, when the time to disruption is also very low. This would raise an alarm, and therefore trigger mitigating actions, when in fact the pulse is non-disruptive. To some extent, such false alarms are difficult to avoid completely, but should be kept to a minimum so as not to jeopardize legitimate experiments.

\begin{figure}[h]
	\centering
	\includegraphics[scale=0.32]{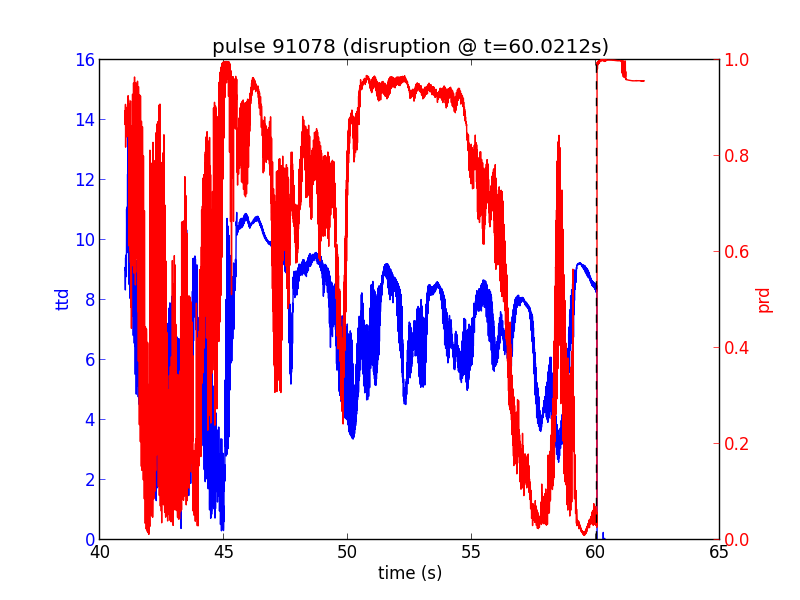}
	\caption{Time to disruption and probability of disruption for pulse 91078.}
	\label{fig_91078}
\end{figure}

Finally, in Figure~\ref{fig_91078} we show an example of the (fortunately) rare situation of missed alarms. At the time of disruption, the probability is low and the time to disruption is still high, which would not raise an alarm. Even if the time to disruption drops to zero and the probability hits 1 at disruption time, this would probably be too late for any mitigating action.

In general, when an alarm is raised and the pulse is disruptive, this is a \emph{true positive} (TP). False alarms correspond to \emph{false positives} (FP) and missed alarms correspond to \emph{false negatives} (FN). When no alarm is raised and the pulse is non-disruptive, it is a \emph{true negative} (TN).

With an upper threshold of 0.85 for the probability of disruption, and a lower threshold of 1.5 seconds for the time to disruption, we can correctly classify (as positive or negative) 85.5\% of the validation pulses. The remaining 14.5\% are divided into false alarms (10.1\%) and missed alarms (4.4\%). It is possible to tweak these thresholds in order to further reduce the missed alarms, but at the cost of increasing the false alarms. It is also possible to tweak the thresholds so that the rate of false alarms and missed alarms is approximately the same.

Table~\ref{tab_thresholds} presents the results when varying the thresholds within reasonable limits. The first row and the first column specify the conditions which will trigger an alarm. It should be noted that only alarms that have been raised up to 30 ms before the disruption are considered, since this is the minimum time to perform mitigation actions at JET~\cite{vries09statistical}. The results in bold (diagonal) illustrate the trade-off between false alarms (FP) and missed alarms (FN).

\begin{table}[h]
	\renewcommand{\arraystretch}{1.1}
	\footnotesize
	\centering
	\begin{tabular}{c|c|c|c|c}
		 & & & & \\
		 & prd $\geq$ 0.95 & prd $\geq$ 0.90 & prd $\geq$ 0.85 & prd $\geq$ 0.80\\
		 & & & & \\
		 \hline
		 ttd $\leq$ 0.5 s & \textbf{\cell{4.4}{81.4}{2.7}{11.5}} & \cell{6.0}{79.5}{4.6}{9.9} & \cell{7.2}{78.2}{5.9}{8.7} & \cell{8.5}{76.9}{7.1}{7.4} \\
		 \hline
		 ttd $\leq$ 1.0 s & \cell{6.8}{81.0}{3.1}{9.1} & \textbf{\cell{9.0}{78.0}{6.1}{6.9}} & \cell{10.3}{75.9}{8.2}{5.6} & \cell{11.0}{74.0}{10.1}{4.9} \\
		 \hline
		 ttd $\leq$ 1.5 s & \cell{8.1}{80.3}{3.8}{7.9} & \cell{10.2}{76.6}{7.4}{5.7} & \textbf{\cell{11.5}{74.0}{10.1}{4.4}} & \cell{12.2}{71.6}{12.4}{3.7} \\
		 \hline
		 ttd $\leq$ 2.0 s & \cell{8.8}{79.5}{4.6}{7.1} & \cell{11.1}{75.5}{8.6}{4.8} & \cell{12.4}{72.3}{11.7}{3.5} & \textbf{\cell{13.3}{69.8}{14.3}{2.7}} \\
	\end{tabular}
	\caption{Prediction results on the validation set (980 pulses) with different alarm-triggering thresholds}
	\label{tab_thresholds}
\end{table}

The results obtained here are comparable to those reported in the literature. For example, the disruption predictor that is currently being used at JET (APODIS) has been reported to have a success rate of 85.38\% and a false alarm rate of 2.46\%~\cite{moreno16disruption}. This seems to be within reach of the results presented in Table~\ref{tab_thresholds}, with a choice of thresholds such as \textit{prd} $\geq$ 0.95 and \textit{ttd} $\leq$ 0.5 s.

Furthermore, when APODIS triggers an alarm for a disruptive pulse, its average warning time is 350 ms in advance of the disruption~\cite{moreno16disruption}. With \textit{prd} $\geq$ 0.95 and \textit{ttd} $\leq$ 0.5 s, we get an average warning time of 929 ms, almost a full second ahead of the disruption.

The average warning time is even higher for every other configuration in Table~\ref{tab_thresholds}, because increasing the \textit{ttd} threshold or decreasing the \textit{prd} threshold has the effect of triggering the alarm sooner. This will, on the other hand, increase the rate of false alarms.

\section{Conclusion}

Deep learning finds several applications in the analysis data coming from fusion diagnostics. Here we have showcased the use of deep neural networks for plasma tomography and disruption prediction. In the first case, a CNN has been trained to replace the computationally expensive task of reconstructing the plasma profile after each experiment. In the second case, a RNN has been trained to detect the onset of plasma disruptions, which can be extremely useful for tokamak operation by raising timely alarms in advance of incoming disruptions.

This means that deep neural networks can be used for both post-processing and real-time processing of fusion data. For real-time processing, fast inference (using special-purpose GPUs, such as the NVIDIA Tesla T4) will be of utmost importance.

\section*{Acknowledgements}
\small
\noindent This work has been carried out within the framework of the EUROfusion Consortium and has received funding from the Euratom research and training programme 2014-2018 under grant agreement No 633053. The views and opinions expressed herein do not necessarily reflect those of the European Commission. IPFN activities received financial support from Funda\c{c}\~{a}o para a Ci\^{e}ncia e Tecnologia (FCT) through project UID/FIS/50010/2013.

\bibliographystyle{elsarticle-num}
\bibliography{paper}

\end{document}